\documentclass[doublecol]{epl2} 
\pdfoutput=1
\usepackage[latin2]{inputenc}
\usepackage{amsmath}
\title{Modularity measure of networks with overlapping communities}

\author{A. L\'az\'ar\inst{1} \and D. \'Abel\inst{1} \and T. Vicsek\inst{1,2}}
\shortauthor{\author{A. L\'az\'ar, D. \'Abel and T. Vicsek}}

\institute{                    
  \inst{1} Department of Biological Physics, Eötvös University - P\'azm\'any P\'eter stny. 1A, Budapest, Hungary H-1117\\
  \inst{2} Statistical and Biological Physics Research Group of HAS - P\'azm\'any P\'eter stny. 1A, Budapest, Hungary H-1117}

\pacs{89.75.Hc}{Networks and genealogical trees}
\pacs{89.75.Fb}{Structures and organization in complex systems}
\pacs{05.90.+m}{Other topics in statistical physics, thermodynamics, and nonlinear dynamical systems}

\abstract{In this paper we introduce a non-fuzzy measure which has been designed to rank the partitions of a network's nodes into overlapping communities. Such a measure can be useful for both quantifying clusters detected by various methods and during finding the overlapping community-structure by optimization methods. The theoretical problem referring to the separation of overlapping modules is discussed, and an example for possible applications is given as well.}

\begin{document}

\maketitle

\section{Introduction}
Networks -- in the sense they are used throughout the present paper -- are basically \emph{graphs} describing real-life complex systems taken from the most different scientific areas, but primarily from biology, economy and sociology. According to recent discoveries, real-life networks tend to have some interesting and rather unexpected common properties, such as scale-free degree distribution, strong disposition to form clusters (also called as communities or modules) or having the so called ``small-world'' property ~\cite{AlbHssz, Nat1, SmllWrld}.\par
Communities (groups of densely interconnected nodes) within these graphs often refer to the \textit{functional units} of the corresponding complex systems, thus their exploration has been a fundamental issue in the study of networks. However, as an important result, these clusters turned out \textit{not} to be separate, but rather overlapping, sharing many edges and nodes.\par
Because of the fundamental role clusters play in real-life networks, many algorithms have been proposed with the aim of uncovering the community-structure of a variety of networks. Earlier ones primarily detect disjoint clusters \cite{NewmanGirvan, NewmanMtx}, meanwhile some of the recent ones detect overlapping modules as well \cite{Nat1, CFinder, LFKertesz}.\par
At the same time, along with the development of the algorithms, arose the demand to define and measure somehow the \emph{``suitability''} of the different partitions provided by the various methods. Moreover, the fact that the concept of ``cluster'' is not specified enough (in the sense that it does not have a widely accepted definition) makes this problem even more ambiguous. However, although some of the proposed \emph{measures} have become widely accepted and used (for example the so called ``Q-modularity'' proposed by Newman and Girvan in \cite{NewmanGirvan}), they are defined only for non-overlapping community structures.\\
\par
Here we would like to note that \emph{fuzzy} measures have been introduced with the same ambition (namely to measure the ``quality'' of an overlapping community-structure, $\left\{ c_1, \ldots, c_K \right\}$) \cite{OlaszAtfedo, FuzzyNepusz} but they share a common constraint: every $i$ node has a ``belonging factor'' $0\leq \alpha_{i,c_r} \leq 1$  which expresses how \emph{strongly} node $i$ belongs to the $r$th cluster $c_r$. The requirement is that 
\begin{equation}
\label{eq.1}
\sum_{r=1}^{K}\alpha_{i,c_r} =1
\end{equation}
for all $i$ belonging to the graph, $K$ denoting the number of clusters.\\
In other words, \emph{none of the nodes can belong to more than one community ``strongly''} (and, primarily, not ``fully''). Recalling social networks, this means that if a person belongs -- let's say -- to her/his family fully (or ``strongly''), then she/he can not belong to other communities, like working place, sport club, etc, only very ``weakly'', or nohow. We believe that this condition is often un-realistic in real-life cases, so our goal has been to define a measure without the above requirement.\\
\par
In brief, the purpose of the present paper is to define a simple but well-usable non-fuzzy measure which, on the one hand, quantifies cluster-structures found by various methods on connected networks, and on the other hand, can be used to detect (overlapping) communities as well by directly optimizing it. For being well-usable, we expect from the measure to take values between -1 and and 1, where a higher positive value corresponds to a better clustering. The zero value expresses random-like network-clustering, and negative values record disadvantageous ones.

\section{The proposed measure}

\begin{figure}
\onefigure[width=8cm]{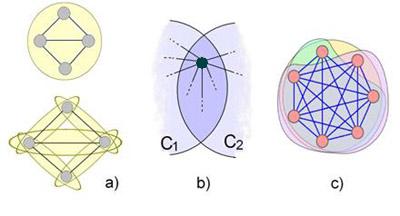}
\caption{\textbf{\emph{a)}} Measure based on the modules \emph{density} will be optimal if all the edges constitute a separate cluster. \textbf{\emph{b)}} An overlapping node that belongs to both the $c_1$ and $c_2$ communities. It contributes with positive values for both clusters. \textbf{\emph{c)}} The appearance of many similar or almost-the-same overlapping communities.}
\label{fig:KisFig1}
\end{figure}

As mentioned above, the notion of ``cluster'' is not well defined: there are many approaches based on different ``intuitive'' characteristics of a community, such as its' denseness, the average path-length among its' nodes, the number of edges going in and out of a given module, the betweenness among nodes belonging to different communities, etc. \cite{SocNtwAB, ClustAnalB, NewmanDet} Although theoretically, measures could be constructed based on any of the above characteristics, in practice, the most commonly used ones exploit the expectation that a cluster should be \emph{``dense''} -- or, as it is often formulated: modules are expected to have relatively more connections within themselves and than among each other ~\cite{NewmanGirvan, NewmanDirGrKit, OlaszAtfedo}. Using the above expectation (clusters should be dense) \emph{and} allowing overlapping community-structure leads to the result that \emph{separate edges} will be returned as optimal  community-structure -- since these are the most dense subgraphs, see fig.~\ref{fig:KisFig1} a. (This happens for example if one tries to apply Newman's Q-modularity directly onto structures where overlapping is enabled.)\\
\par

According to our experiments, \emph{none} of the ``intuitive approaches'' is enough to create a suitable measure \emph{alone}, because they result in ``degenerated structures'' to be optimal ones, similar to the one seen above. On the other hand, \emph{combinations} of approaches can handle this phenomenon.\\
We have obtained good results by utilizing the following expectations: (1) the edges of a given node should primarily go inward its' cluster(s) and should not go outward, and (2): clusters should be dense. The first criterion shows how ``justifiable'' it is to assign the node $i (\in c_r)$ to the $r$th cluster $c_r$: it is the difference between the \emph{inward} going edges ($\sum_{j\in c_r, i\neq j}a_{i j}$) and the \emph{outward} going edges ($\sum_{j\notin c_r}a_{i j}$), divided by the $d_i$ degree of node $i$. Put it together, we get that every $i$ node contributes to the $r$th cluster to which it belongs to with the following value:

\begin{equation}
\label{eq.2}
\frac{\sum\limits_{j\in c_r, i\neq j}a_{i j} - \sum\limits_{j \notin c_r}a_{i j}} {d_i}
\end{equation}

where $a_{i,j}$ denotes the proper element of the adjacency matrix defining the network, interpreted as usually, that is, 

\begin{equation}
\label{eq.3}
a_{ij} = \left\{
	\begin{array}{ll}
		1 & \mbox{if $i$ and $j$ are connected,}\\
		0 & \mbox{if not}
	\end{array}	
\right.
\end{equation}

The more edges go inward and the less edges go outward the cluster, the more the above ratio converges to 1. If more edges go outward than inward, the expression is negative, and if all of them go outward, the result is -1. Since a node can contribute with positive values to more than one clusters -- due to the overlapping areas, see fig.~\ref{fig:KisFig1} b -- the whole network's modularity value is higher if a node like that belongs to both modules.\par
To avoid community-structures having only a few communities with very high $M^{ov}_{c_r}$ values, we add the criterion that all nodes have to belong to at least one module. (A trivial solution for that is, to put all the left-out nodes into a separate cluster at the end. We have obtained our results like this too.)
Also, the appearance of many similar or almost-the-same overlapping communities (as it can be seen on fig. \ref{fig:KisFig1} c) is avoidable by dividing the above expression by the number of clusters $i$ belongs to, denoted by $s_i$. Thus the $r$th cluster, $c_r$ will contribute to the final result $M^{ov}$ with:

\begin{equation}
\label{eq.4}
M_{c_r}^{ov} = \frac{1}{n_{c_r}} \sum\limits_{i \in c_r} \frac{\sum\limits_{j\in c_r, i\neq j}a_{i j} - \sum\limits_{j \notin c_r}a_{i j}} {d_i \cdot s_i}   \cdot \frac{n^e_{c_r}}{\binom{n_{c_r}}{2}}
\end{equation}

where $n_{c_r}$ is the number of nodes and $n^e_{c_r}$ is the number of edges that the $r$th cluster $c_r$ contains, respectively.\\

The \emph{density} of a module -- which was our ``second requirement'' -- is straightforward to interpret as $\frac{n^e_{c_r}}{\binom{n_{c_r}}{2}}$. This expression gives 1 if the $r$th module $c_r$ (which is a (sub)graph) contains all its' possible edges, and 0 if it does not have any of them. Since the first factor ranges between -1 and 1, the second factor between 0 and 1, the whole expression varies between -1 and 1.\\

This remains true for the final measure $M^{ov}$ as well, which is the average of the $M_{c_r}^{ov}$ module-values:\\ 
$M^{ov} = \frac{1}{K} \sum\limits_{r=1}^K M_{c_r}^{ov}$, that is,

\begin{equation}
\label{eq.5}
M^{ov} = \frac{1}{K} \sum\limits_{r=1}^K \left[ \frac{ \sum\limits_{i \in c_r} \frac{\sum\limits_{j\in c_r, i\neq j}a_{i j} - \sum\limits_{j \notin c_r}a_{i j}} {d_i \cdot s_i}  }{n_{c_r}} \cdot \frac{n^e_{c_r}}{\binom{n_{c_r}}{2}} \right]  
\end{equation}
\par

Since the density of clusters containing one single node (when $n_{c_r}=1$) is not defined (because $\binom{1}{2}$ is not defined), we simply set their $M_{c_r}^{ov}$ modularity value to zero. (Isolated nodes (when $d=0$) can not appear, since the network assumed to be connected.)\\
\par
Here we would like to note that handling the unclustered nodes (nodes that do not belong to any of the modules) is possible in many ways. We have chosen to put them into a separate community, but some kind of \emph{weighting} is also conceivable, when the weight is in inverse proportion to the number of the unclustered nodes (the more nodes are clustered, the higher the final score is). Furthermore, one can consider the \emph{weighting} of the clusters according to their sizes as well.

\section{One cluster or more clusters? -- When to separate?\nodot}
\begin{figure}
\onefigure[width=8cm]{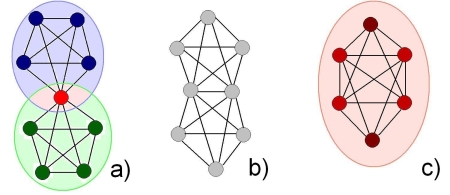}
\caption{The question when to handle a (sub)graph as one community and when as more, is non-trivial, because ``intuition'' gives different answers to different people. At the same time, most of us would agree on separating two 5-cliques overlapping in one single node \emph{\textbf{(a)}}, but handling them as one community, if they share 4 nodes \textbf{\emph{(c)}}. Cases between \textbf{\emph{(b)}} are a matter of ``taste''.}
\label{fig:WhenToSeparate}
\end{figure}

This question is highly non-trivial, because it is -- up to a great extent -- simply a matter of ``intuition'' or taste, being different from person to person. For example most of us would agree on separating two 5-cliques overlapping in one single node, but handling them as one community, if they share 4 nodes (see fig.~\ref{fig:WhenToSeparate}). But what is the case, if they share two or three nodes?\\

\begin{figure}
\onefigure[width=8cm]{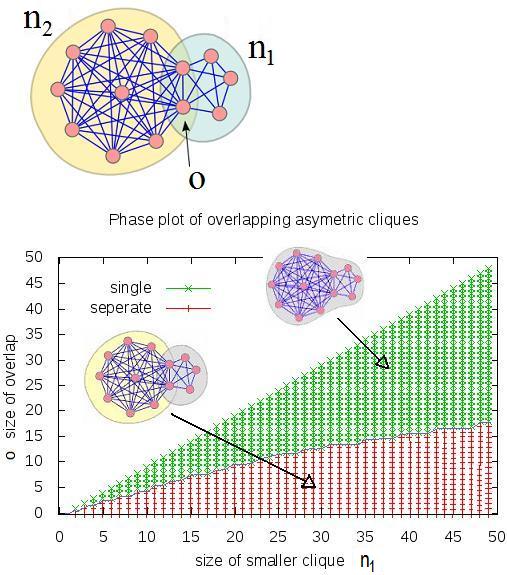}
\caption{Given a complete-graph with $n_2=50$ nodes and a smaller one with $n_1$ nodes ($n_1 \in \left\{1 \ldots 50 \right\}$, also complete-graph; $n_1$ is shown on the horizontal axis). These two graphs overlap in $o$ nodes (vertical axis), where $o \in \left\{1 \ldots n_1 \right\}$. The $n_1-o$ parameter-pairs generate two dissevering regions: the upper one is where the introduced measure, $M^{ov}$ gives higher score if the two graphs are handled as one module, while the lower one covers those $n_1-o$ pairs, which give higher score if the graphs make up separate communities.}
\label{fig:PhasPlotDiag}
\end{figure}

Figure~\ref{fig:PhasPlotDiag} describes how the introduced measure, $M^{ov}$ behaves with respect to the above question. Given a complete-graph with $n_2=50$ nodes and a smaller one with $n_1$ nodes ($n_1 \in \left\{1 \ldots 50 \right\}$, also complete-graph). These two graphs overlap in $o$ nodes, where 
$o \in \left\{1 \ldots n_1 \right\}$. The horizontal axis shows the size of the smaller graph, $n_1$, while the vertical axis shows the number of the overlapping nodes ($o$) between the two graphs. Two regions show up: 
the lower region covers the $o-n_1$ parameter-pairs by which $M^{ov}$ gives higher score if the two graphs are handled as separate communities, while the upper one covers those $n_1$-$o$ pairs, which give higher score, if the overlapping graphs are handled as one module. One extreme is when the overlap is 0 (the two graphs do not share any nodes, horizontal axis) -- which obviously falls in the lower, ``separate''-region. The other end-value is when they share all the $n_1$ nodes, that is, the smaller graph (the $n_1$-clique) is a real sub-graph, a part of the bigger complete-graph -- this case is represented by the diagonal line starting from the pole.

\section{An application}

\emph{CFinder}, an algorithm designed to uncover the overlapping community-structure of networks \cite{Nat1}, has a ``tuning-parameter'' ($k$) which determines the \emph{cohesiveness} of the revealed modules: the higher the parameter $k$, the smaller, the more disintegrated, but at the same time the more cohesive are the detected communities. This is a result of the method, which exploits the observation, that a typical community consists of several complete subgraphs that tend to share many of their nodes. The algorithm uncovers those modules which form so called ``$k$-clique communities'', that is, unions of $k$-cliques that can be reached from each other through a series of adjacent $k$-cliques.\\
Theoretically $k$ can be any positive integer starting from 3, but in practice it is usually smaller than ten. (If $k=2$, CFinder detects the connected subgraphs, that is, those modules which are unions of 2-cliques (which are edges) and can be reached from each other through a series of adjacent edges.)
The proper value of $k$ depends on the network. In the following we define the most proper $k$ for some real-life networks using the introduced measure, $M^{ov}$.\\
 
\begin{figure}
\onefigure[width=8cm]{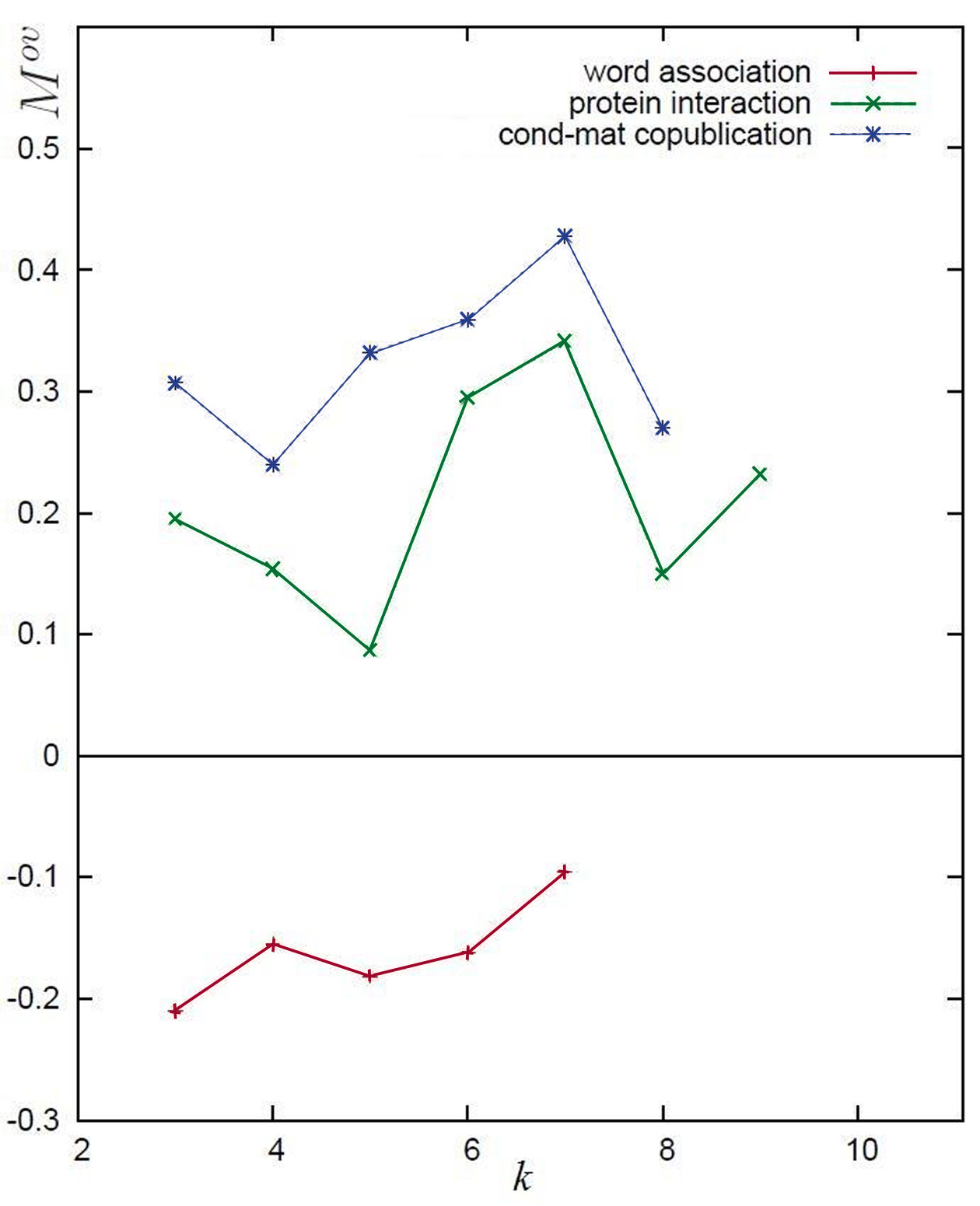}
\caption{The $M^{ov}$ scores as a function of the $k$ ``tuning-parameter'' belonging to the CFinder algorithm, for three real-life networks: (1) cond-mat publication (topmost curve) (2) protein interaction and (3) word association (bottommost curve). The suggested $k$-values are those where the curves reach their maximum.}
\label{fig:CPM}
\end{figure}

Figure~\ref{fig:CPM} depicts the $M^{ov}$ scores as a function of the $k$ parameter for three real-life networks: (1) word association, (2) protein interaction, and (3) cond-mat publication. \par

The nodes of the first graph, `word association', are words which are linked if the people in a survey associated them with each other \cite{DBaseWAssoc}. (Originally it is a weighted, directed graph, where the weight of an edge indicates the frequency that the people associated the end point of the link with its' start point, but here we have used a simplified -- undirected, unweighted -- form of it.)
The `protein interaction' network describes the protein-protein interactions in S. cerevisiae (see details in \cite{DBasePPI}), and finally, the `cond-mat publication' network describes co-authorships among mathematicians, obtained from the Los Alamos cond-mat archive (\cite{CondMatWeb}). (Originally this is a weighted graph as well, where the weights are proportional to the number of common works, but, here too, we have used a simplified, unweighted version of the graph, in which the edges have been eliminated under a certain threshold-weight. See more details in \cite{DBaseCondMat}.)\par

As it can be seen on fig.~\ref{fig:CPM}, in the case of the protein-interaction network and the cond-mat publication, both curves reach their maximum at $k=7$, which is their optimum value for $k$.

The word-association network displays a very interesting behavior: the whole curve is in the negative region. This is most probably due to the fact that this graph contains many words with several meanings, e.g., the word ``bright'', which -- according to the survey -- is often associated with words having alternative meanings, like ``smart'', ``light'', ``dark'', ``sun'', etc. Accordingly, in a graph like this, if slightly overlapping modules arise around the different meanings of a word, and if between the nodes of these otherwise separate modules there are relatively many edges (associations) a negative numerator in $M^{ov}$ is resulted.\\
\par

\acknowledgments
This research was partially supported by the following grant: EU ERC COLLMOT.\\
\par


\end{document}